\documentstyle[twocolumn,aps,epsfig,floats]{revtex}
\begin{document}

\def\DESepsf(#1 width #2){\epsfxsize=#2 \epsfbox{#1}}
\def\Bb{\textsf{\textbf{B}}}
\def\Dd{\textsf{\textbf{D}}}

\newcommand{\beq}{\begin{equation}}
\newcommand{\eeq}{\end{equation}}
\newcommand{\beqa}{\begin{eqnarray}}
\newcommand{\eeqa}{\end{eqnarray}}
\newcommand{\sss}{\scriptscriptstyle}

\newcommand{\dida}[1]{/ \!\!\! #1}
\renewcommand{\Im}{\mbox{\sl{Im}}}
\renewcommand{\Re}{\mbox{\sl{Re}}}
\def\simge{\hspace*{0.2em}\raisebox{0.5ex}{$>$}
     \hspace{-0.8em}\raisebox{-0.3em}{$\sim$}\hspace*{0.2em}}
\def\simle{\hspace*{0.2em}\raisebox{0.5ex}{$<$}
     \hspace{-0.8em}\raisebox{-0.3em}{$\sim$}\hspace*{0.2em}}

\def\vday{{20/11}}
\def\check{$\Rightarrow$\marginpar{$\Leftarrow$\tiny{\vday}} }

\twocolumn[\hsize\textwidth\columnwidth\hsize\csname 
@twocolumnfalse\endcsname
\title{\boldmath{
$B$ meson decays to baryons in the diquark model}
      }
\vfill
\author{$^{1}$Chia-Hung V. Chang and $^2$Wei-Shu Hou}
\address{
\rm $^1$Department of Physics, National Taiwan Normal University,
Taipei, Taiwan, R.O.C.\\
\rm $^2$Department of Physics, National Taiwan University,
Taipei, Taiwan, R.O.C.
}

%
%
\vfill
\maketitle
\begin{abstract}
We study $B$ meson decays to two charmless baryons in the diquark model,
including strong and electroweak penguins as well as the tree operators.  
It is shown that penguin operators can enhance 
$\bar{B} \rightarrow \Bb_s \bar{\Bb}$ considerably, 
but affect $\bar{B} \rightarrow \Bb_1 \bar{\Bb}_2$ only slightly, where 
$\Bb_{(1,2)}$ and $\Bb_s$ are non-strange and strange baryons, respectively.
The $\gamma$ dependence of the decay rates due to 
tree-penguin interference is illustrated.
In principle, some of the $\Bb_s \bar{\Bb}$ modes could dominate
over $\Bb_1 \bar{\Bb}_2$ for $\gamma > 90^\circ$,
but in general the effect is milder than their mesonic counterparts.
This is because the $O_6$ operator can only produce vector 
but not scalar diquarks, while the opposite is true for $O_1$ and $O_4$.
Predictions from diquark model are compared to those from the sum
rule calculation. The decays $\bar{B} \rightarrow \Bb_s \bar{\Bb}_s$ and
inclusive baryonic decays are also discussed. 

\end{abstract}
\vskip2pc]

%

\pagestyle{plain}

\section{Introduction}
$B$ meson decays provide a unique setting for baryon pair production,
since it is impossible for the $D$ system. 
Once observed, these decays could shed light on 
our understanding of baryon production,
and may offer further probes \cite{housoni} of 
underlying weak decay dynamics such as $CP$ violating phases.  

Many rare mesonic $B$ decays have been observed 
at the $10^{-5}$ level in recent years, 
heralding the start of the ``B Factory" era.
However, rare baryonic decays have yet to be discovered. 
The most recent published limits come from 
the CLEO collaboration \cite{CLEO}.
Based on 5.8 million $B\bar B$ events,
CLEO finds $B \rightarrow \bar{\Lambda}p, \, \bar{\Lambda}p\pi^-$ and 
$p\bar{p} < 0.26,\, 1.3,$ and $0.7\times 10^{-5}$, respectively.
There was some $2.8 \sigma$ excess in the 
$\bar{B}^0 \rightarrow p\bar{p}$ channel, 
but it was insufficient to claim discovery.
The B factories, i.e. the Belle and BaBar Collaborations,
have now each accumulated an order of magnitude more data.
A preliminary result from Belle \cite{ppbar} has pushed the 
$\bar{B}^0 \rightarrow p\bar{p}$ limit down to the $10^{-6}$ level. 
This rules out the CLEO hint, and 
puts two-body baryonic modes in strong contrast to
the corresponding mesonic modes.  
However, though still elusive, it is quite possible that 
charmless baryonic modes are just around the corner. 

Theoretical work on rare baryonic decays is sparse. 
Most of them were stimulated by the surprising 
(and false \cite{bebek}) 1987 results \cite{argus} of
Br$(B^- \to p\bar{p}\pi^-) = (3.7 \pm 1.3 \pm 1.4)\times 10^{-4}$ and 
Br$(\bar B^0 \to p\bar{p}\pi^+\pi^-) = (6.0 \pm 1.3 \pm 1.4)\times 10^{-4}$
from the ARGUS Collaboration. 
Pole models \cite{soni,jarfi} and the sum rule approach \cite{sumrule} 
have been proposed for calculating the two body decay widths,
while Ball and Dosch \cite{ball} pursued the diquark model appoach. 
All of these works are now a decade old. 
Not until very recently, 
sensing that experiment is about to move forward, 
did theorists start to pay attention again. 
Hou and Soni \cite{housoni} pointed out 
the need for reduced energy release on the baryon side,
e.g. charmless baryonic $B$ decays may be more prominent
in association with $\eta'$ or $\gamma$.
Chua, Hou and Tsai studied $\bar{B} \rightarrow D^{*-} N
\bar{N}$ \cite{three1} and $B^0 \rightarrow \rho^- p \bar{n}, \pi^- p
\bar{n}$ \cite{three2} using a factorization approach
of current produced baryons.

For two body baryonic decays, the sum rule and diquark model approaches 
are the most relevant. 
The sum rule calculation \cite{sumrule} predicts that 
the branching ratio of $\bar{B} \rightarrow \Bb_1 \bar{\Bb}_2$ is 
typically of $10^{-6}$ order while 
Br$(\bar{B} \rightarrow \Bb_s \bar{\Bb})  \approx 
(0.3 \sim 1.0) \times 10^{-5}$ and  
Br$(\bar{B} \rightarrow {\rm \bf B_c \bar{B}_{(c)}}) \approx 
{\rm O}(10^{-3})$.  
Here $\Bb_{(1,2)}$, $\Bb_s$ and $\Bb_c$ denote non-strange,
strange and charmed baryons, respectively.
The CLEO limit of 
Br$(\bar{B}^0\rightarrow p\bar{\Lambda})< 0.26 \times 10^{-5}$ \cite{CLEO}
is already at odds with the sum rule prediction of 
Br$(B^+\rightarrow p\bar{\Lambda}) \sim 1.0 \times 10^{-5}$.
Furthermore the predicted 
Br$(\bar{B}^0 \rightarrow p\bar{p})\sim 1.0 \times 10^{-6}$ is
an order of magnitude smaller than 
$B^-\rightarrow p\bar{\Lambda}$.  
This is in contrast with the diquark model \cite{ball} which 
typically gives a larger rate for 
$\bar{B} \rightarrow \Bb_1 \bar{\Bb}_2$ over $\Bb_s \bar{\Bb}$,
although it can predict really only relative rates.  
For example, the $\bar{B}^0 \rightarrow p\bar{p}$
mode has larger rate than all the 
$\bar{B} \rightarrow \Bb_s \bar{\Bb}$ decays.  
Unfortunately, Ref. \cite{ball} did not include penguin operators.
Judging from the role played by penguins in mesonic $B$ decays like
$B\rightarrow \pi \pi, \, K\pi$,  
an enhancement by penguins to ${\rm Br}(\bar{B} \rightarrow \Bb_s \bar{\Bb})$ 
is to be expected. Furthermore, with penguins ignored, 
${\rm Br}(B^- \rightarrow\Lambda \bar{p})=0$ in the diquark model,
preventing us from comparing with Belle and CLEO search directly.
 
The purpose of this paper is to complete the diquark model treatment of 
two body charmless baryonic $B$ decays by including penguin diagrams,
and to assess the importance of these penguin effects.
Whether the approach of using diquarks to describe baryon formation 
is correct or not is still an open question. 
Even if one assumes the idea is reasonable, 
to calculate the relative baryonic decay rates, 
we still need to make many other dynamical assumptions, 
which introduce further uncertainties.
Our goal in this paper is simply to clarify the actual predictions of 
the diquark model by expanding on the work of Ref.\ \cite{ball}.
The experimental measurements in the future will decide 
whether diquark model or sum rule approach is more 
relevant in the description of baryonic decays,
or how they might be improved upon.  
Thus, 
we do not attempt at improving the diquark model 
towards absolute rate calculations.

We find that penguin operators indeed could 
enhance $\bar{B} \rightarrow \Bb_s \bar{\Bb}$ decay rate
by a factor of $\sim 5$ for $\gamma = 90^{\circ}$. 
Due to tree-penguin interference, the decay widths
now depend on the unitarity phase angle 
$\gamma \,\,(\equiv {\rm arg} \, V^*_{ub}$, in the convention
of PDG \cite{PDG}). 
The enhancement in baryonic $B$ decays is milder than in the mesonic
decays because the operator $O_6$ can not generate scalar diquarks. 
Penguins also affect non-strange decays, 
$\bar{B} \rightarrow \Bb_1 \bar{\Bb}_2$, 
introducing also a $\gamma$ dependence, but the effect here is small.
In general, $\bar{B} \rightarrow \Bb_s \bar{\Bb}$ is still smaller 
than $\bar{B} \rightarrow \Bb_1 \bar{\Bb}_2$, and 
$\bar{B}^0 \rightarrow p \bar{p}$ typically has the largest rate.
But for large $\gamma > 90^\circ$,
the $\bar{B} \rightarrow \Sigma^+ \bar{p}, \,
\Sigma^+ \overline{\Delta^{++}}$ rate could become
larger than $\bar{B}^0 \rightarrow p \bar{p}$.
We find that the pattern of decay widths calculated using diquark model 
is quite different from the sum rule results. 
More experimental data should shed light on the two models.

This paper is organized as follows. 
We first review the diquark model and baryonic $B$ decays.
The connection between penguin and diquark operators is discussed.
In Sec.~III, we study inclusive baryonic decays,
with two body exclusive decays discussed in Sec. IV, 
in both cases including the effect of penguins. 
The conclusion is given in the last section.

\section{Diquark Model and Penguin Operators} 

It is well known that the strong force between 
two quarks in a color-antitriplet combination is attractive, 
hence  it has been speculated for a long time that they will 
form a bound or correlated state, called the diquark.
The flavor antisymmetric combinations 
form scalar diquarks while
flavor symmetric combinations
form vector diquarks. 
The diquark picture is useful in the description of baryons.
The spin $1/2$ octet and spin $3/2$ decuplet baryons can be understood as 
bound states of a quark and a scalar or vector diquark, respectively
\cite{diquarksum}. 

Diquarks can be generated in weak decays \cite{ionehalf,NY,NSPRD}.
The tree level weak decay effective Hamiltonian is
\begin{eqnarray}
{\cal H}_{\rm eff}&=&{G_F\over\sqrt{2}} \sum_{i=u,c}
\Bigg\{ V_{iq}^*V_{ib}\Big[c_1(\mu)O_1^i(\mu)+c_2(\mu)O_2^i(\mu)\Big] \Bigg\}
\nonumber \\
    &  &    +{\rm h.c.},
\end{eqnarray}
where
\begin{eqnarray}
 O_1^u &  = & (\bar q_\beta u_\alpha)_{V-A}(\bar u_\alpha b_\beta)_{V-A},
    \nonumber \\
 O_2^u &  = & (\bar qu)_{V-A}(\bar ub)_{V-A},
    \nonumber 
\end{eqnarray}
and likewise for $O_{1,2}^c$. The quark $q$ could be $d$ or $s$.
The current-current operators are defined as
 $(\bar q_1q_2)_{_{V- A}}\equiv\bar q_1\gamma_\mu(1- \gamma_5)q_2$.
The operators could be rewritten, after a Fierz transformation, 
in terms of scalar diquark field operators.
For example:
\beqa
 [c_1 O_1^u+c_2 O_2^u ]
& = & - (c_2-c_1) 
   \, (ub)_{Lk}
   (ud)_{Lk}^{*}  \nonumber 
\\
& & +{\rm \, color  \,\,sextet \,\, current}, \label{Fierz}
\eeqa
with the scalar diquark field operator defined as
\beq
(qQ)_{Lk}=\epsilon_{klm} (\bar{q}^{\scriptscriptstyle C}_l (1-\gamma_5) Q_m),
\eeq
where $k,l,m$ are color indices.
Here $\bar{q}^{\scriptscriptstyle C} \equiv q^T C$.
The scalar diquark field operator can create a scalar diquark 
from vacuum with the strength $g_{qQ}$, 
usually called ``diquark decay constant": 
\beq
\langle 0|\epsilon_{klm}(\bar{q}^{\scriptscriptstyle C}_l \,\gamma_5\, Q_m)
|(qQ)_k^{0+}\rangle \equiv \sqrt{\frac{2}{3}} \delta_{il} g_{qQ}.
\label{decaycons}
\eeq

We see from Eq.\ (\ref{Fierz}) that
a $b$ quark can decay into a scalar diquark plus an antiquark. 
Since the baryons are bound states of a diquark and a quark, 
it is natural to expect that in decays to baryonic final states,
the diquark operators will dominate and 
the sextet current operators can be ignored.
Following this reasoning, Ref.\ \cite{ball} gives a picture 
of two body baryonic $B$ decays.
The antiquark produced in the decay combines with 
the spectator antiquark to form a scalar or vector anti-diquark.
As the diquark and anti-diquark fly apart, they
pull a  quark-antiquark pair out from the vacuum, 
resulting finally in a baryon-antibaryon pair. 
Of course, this may not be the only mechanism, 
but it is assumed to be the dominant one
in the diquark model \cite{ball}. 
It is interesting to note that $O_{1,2}$ can only generate scalar diquarks, 
hence $\bar{B}$ decaying to a decuplet baryon plus either an octet or
decuplet antibaryon, $\bar{B} \to \Bb^* \overline{\Bb}^{(*)}$,
are predicted to have small rates \cite{ball}. 
These decays can only arise from the penguin operators $O_{5,6}$,
as will be discussed below.


In addition to the tree level effective Hamiltonian, 
it is well known that penguin diagrams are important for 
the charmless decays of $B$ mesons. 
In mesonic decays like $B \rightarrow K\pi$ and $\pi\pi$, penguin diagrams
are crucial in the calculation of decay rates and CP asymmetries. 
Their effects can be  
described  by the effective penguin operators $O_3$ through $O_{10}$,
\begin{equation}
{\cal H}_{\rm penguin}= -{G_F\over\sqrt{2}}
\Bigg\{ 
       V_{tq}^*V_{tb}\sum^{10}_{i=3}c_i(\mu)O_i(\mu)\Bigg\}+{\rm h.c.},
\end{equation}
where 
\begin{eqnarray}
   O_{3(5)} &  = & \sum_{q'}(\bar q'q')_{{V-A}({V+A})}(\bar qb)_{V-A},
 \nonumber \\
   O_{4(6)} &  = &\sum_{q'}(\bar q'_\beta q'_\alpha)_{{V-A}({V+A})}
            (\bar q_\alpha b_\beta)_{V-A},\nonumber \\
  O_{7(9)}&  = &{3\over 2}\sum_{q'}e_{q'}(\bar q'q')_{{V+A}({V-A})}
            (\bar qb)_{V-A},  \nonumber \\
   O_{8(10)} &  = &{3\over 2}\sum_{q'}e_{q'}
             (\bar q'_\beta q'_\alpha)_{{V+A}({V-A})}
             (\bar q_\alpha b_\beta)_{V-A},
\end{eqnarray}
with $O_{3-6}$, $O_{7-10}$ the QCD and electroweak penguin operators,
respectively, and $q=d,\;s$,
$(\bar q_1q_2)_{_{V\pm A}}\equiv\bar q_1\gamma_\mu(1\pm \gamma_5)q_2$.
The sum over $q^\prime$ runs over all quark flavors 
that exist in the effective field theory.

The penguin operators can also be written in terms of diquark operators.
Let us consider $b\to s$ penguins
(similar discussion follow for the $q=d$ case).
The $(V-A)\times(V-A)$ type penguin operators 
$O_{3,4}$ and $O_{9,10}$ are similar in form to $O_{1,2}$.
They can be written, after a  Fierz transformation,
as the sum of diquark operators $\pm\sum_{q'} (q'b)_L (q's)_L^{\dagger}$ 
plus color sextet terms. Again the latter will be ignored.
The sum includes the operator $(ub)_L \, (us)_L^{\dagger}$, 
which is of exactly the same form as that obtained from $O_{1,2}$, 
as well as $(db)_L \, (ds)_L^{\dagger}$, 
which is not present in the tree operators. 
Note that the $q'=s$ piece $(sb)_L \, (ss)_L^{\dagger}$ vanishes 
since the 
scalar diquark is antisymmetric in their flavor 
constituents. 

On the other hand, the $(V-A)\times(V+A)$ type penguin operators 
$O_{5}$ and $O_6$ do not give rise to scalar diquark operators.
The reason is that a scalar diquark is a quark-quark correlation. 
The operators $O_{5,6}$ contains 
one left-handed and one right-handed quark field,
which can not form a scalar combination under the Lorentz transformation.
In other words, the scalar diquark content of octet baryons 
implies that the operators $O_{5,6}$ 
(and the corresponding electroweak penguin operators $O_{7,8}$)
do not contribute to the $\bar{B}$ decays to octet baryon 
plus either an octet or decuplet antibaryon: 
$B \to \Bb \overline{\Bb}^{(*)}$.
This is contrary to a significant role that they play
in the mesonic decays such as $B \rightarrow K\pi, \, \pi\pi$. 
However, $O_{5,6}$ could generate operators 
that consist of vector diquarks
which $O_{1}$ through $O_{4}$ could not produce. 
By Fierz transformation, for example:
\beqa
 && c_5  (\bar{s} \gamma^{\mu} b)_{\sss V-A}
          (\bar{q'} \gamma_{\mu} q')_{\sss V+A} +
  c_6 (\bar{s}_{\alpha} \gamma^{\mu} b_{\beta})_{\sss V-A}
          (\bar{q'}_{\beta} \gamma_{\mu} q'_{\alpha})_{\sss V+A} \nonumber \\
 && = - \frac{1}{2} (c_6-c_5) 
   \,(q'_{R} b_{L})^*_k     (q'_{R} s_{L})^{*\dagger}_k  
 +{\rm color\ sextet},
\eeqa
with the vector diquark field operator defined as
\beq
(q_{R} Q_{L})^*_k \equiv
\epsilon_{klm} (\bar{q}^{\scriptscriptstyle C}_l \gamma^{\mu}(1-\gamma_5) Q_m).
\eeq
Since decuplet baryons are bound states of a vector diquark and a quark,
$O_{5,6}$ will produce $\bar{B}$ decays to 
a decuplet baryon plus either an octet or decuplet
antibaryon: $ \bar{B} \to \Bb^* \overline{\Bb}^{(*)}$. 
As mentioned above, these decays can not be generated 
by the tree $O_{1,2}$ and the penguin $O_{3,4}$ operators.    
For example, the novel channel  $B^- \rightarrow \Omega^- \overline{\Xi}^0$ 
(five strange quarks in final state) 
could arise from $B^- \rightarrow  (ss)^*(\overline{su})$.

To sum it up, the effective Hamiltonian that generates scalar diquarks 
can now be collected as
\begin{eqnarray}
{\cal H}_{\rm diquark}&\sim& -{G_F\over\sqrt{2}}
\bigg\{{\cal A}_1
            (ub)_L \,    (ud)_L^{\dagger}   +
        {\cal A}_2
          (ub)_L \, (us)_L^{\dagger}  \nonumber \\
& & + {\cal A}_3 (sb)_L \, (sd)_L^{\dagger}
           +{\cal A}_4 (db)_L \, (ds)_L^{\dagger}
          +{\rm h.c.} \bigg\}, \nonumber
\end{eqnarray}
with the coefficients   
\begin{eqnarray}
{\cal A}_1 & \equiv & V_{ud}^*V_{ub}(c_2-c_1)-V_{td}^*V_{tb}(c_4-c_3+c_9-c_{10}),
		\nonumber \\
{\cal A}_2 & \equiv & V_{us}^*V_{ub}(c_2-c_1)-V_{ts}^*V_{tb}(c_4-c_3+c_9-c_{10}),
		\nonumber \\
{\cal A}_3 & \equiv & -V_{td}^*V_{tb}(c_4-c_3-\frac{1}{2} c_9+\frac{1}{2}c_{10}),
		\nonumber \\
{\cal A}_4 & \equiv & -V_{ts}^*V_{tb}(c_4-c_3-\frac{1}{2} c_9+\frac{1}{2}c_{10}).
\label{a1a2}
\end{eqnarray}
For vector diquarks, we have
\begin{eqnarray}
{\cal H}_{\rm diquark*}&\sim& -{G_F\over\sqrt{2}}
\sum_{q'=u,d,s} \bigg\{ {\cal B}_1 (q'_{R} b_{L})^* (q'_{R} d_{L})^{*\dagger}
   \nonumber \\
 && + {\cal B}_2 (q'_{R}b_{L})^*(q'_{R}s_{L})^{*\dagger} + {\rm h.c.} \bigg\},
\end{eqnarray}
with the coefficients   
\begin{eqnarray}
{\cal B}_1 &\equiv& -\frac{1}{2} V_{td}^*V_{tb}
                   (c_5-c_6+ \frac{3}{2} e_{q'} c_7-\frac{3}{2} e_{q'} c_{8}),
               \nonumber \\
{\cal B}_2  &\equiv& -\frac{1}{2} V_{ts}^*V_{tb}
                   (c_5-c_6+ \frac{3}{2} e_{q'} c_7-\frac{3}{2} e_{q'} c_{8}).
\end{eqnarray}

\section{Inclusive Baryonic Decays}

Before we discuss the more difficult exclusive baryonic decays, 
the diquark picture could actually
give us useful insight on the inclusive baryonic $B$ decays.

In the diquark model, one postulates that baryons are bound states of 
a (scalar or vector) diquark and an antiquark.
It is natural to expect that baryonic $B$ decays  
proceed dominantly via the process of 
$b$ quark decaying into a scalar diquark plus an anti-quark. 
Since the subsequent hadronization process always generate at least one baryon,
with the other antibaryon guaranteed by baryon number conservation, 
the inclusive baryonic decay rates can be approximated by 
the rates of $b \rightarrow \Dd \bar{q}$ (Fig.\ 1), 
where $\Dd$ denotes 
a scalar diquark such as $(ud)$ or $(cd)$.

\begin{figure}[t!]
\centerline{\DESepsf(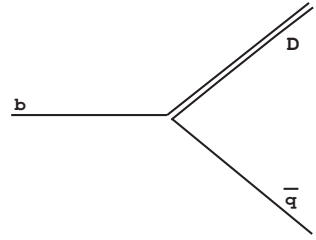 width 4.7cm)
}
\caption{
The inclusive decay $b \rightarrow \Dd \bar{q}$.
}
\end{figure}  

This is the approach mentioned by Neubert and Stech 
in Refs. \cite{NY} and \cite{stech}.
Though citing the result from the above two papers, Ref.\ \cite{ball}
adopted a different method to calculate the inclusive rates. 
It computes the rates of the
$B$ meson decaying into a diquark and an anti-diquark, 
i.e. $B \rightarrow \Dd \overline{\Dd'}$. 
The two body channel $B \rightarrow \Dd \overline{\Dd'}$, 
in which both diquarks 
are fast moving, implies that at least two fast moving baryons 
will be generated. 
Though baryons always appear in pairs in baryonic decays, 
this could be too restrictive for inclusive baryonic decays,
since it ignores the possibility of slow moving antibaryons.

The $b \to \Dd \bar{q}$ rate turns out to be rather close to 
the observed  inclusive rate. Let us take a closer look.
After the $b \to \Dd \bar{q}$ decay, the diquark $\Dd$, 
the antiquark $\bar{q}$ and the spectator antiquark 
jointly form a color singlet, just like the three antiquarks in an antibaryon.
The fast moving diquark $\Dd$ pulls out a quark $q^\prime$ from vacuum
to form a baryon, leaving behind a slow antiquark $\bar{q}^\prime$. 
The color configuration of $\bar{q}^\prime$, $\bar{q}$ 
and the spectator antiquark is again just like that of an antibaryon. 
Since $\bar{q}$ is moving fast while the other two are slow, 
the system breaks up into hadrons through fragmentation.  
This generates all kinds of possible final products, but at least 
one antibaryon has to be generated due to baryon number conservation.
One possible scenario is for $\bar{q}$ to form a fast moving antibaryon
by pulling the two antiquarks with it.
Two body baryonic decays are just such a case.
Another scenario is that the fast moving $\bar{q}$ 
captures one quark to form a meson, leaving behind a slower antiquark. 
The final products of the decay then consist of a fast baryon, a fast meson
and a slow antibaryon plus possible soft mesons. 
The baryon-antibaryon pair mass would then be far below $m_{B}$.
One could also break two strings and capture two new antiquarks to form a fast
antibaryon with the remaining quarks and antiquarks 
combining into mesons.
The final products would then be one fast moving baryon, one fast antibaryon,
plus two (or more) soft mesons.  
In all the above scenarios, one baryon and one antibaryon are generated.
But the second scenario clearly is not included  
in the $B \rightarrow \Dd \overline{\Dd'}$ picture of Ref.\ \cite{ball}.

We list the branching ratios of $b \rightarrow \Dd \bar{q}$ decays in Table I.
The transition amplitude of $b \rightarrow \Dd \bar{q}$ is assumed to 
factorize into the product of the diquark decay constant 
as defined in Eq.\ (\ref{decaycons}), 
and the quark level amplitude $\langle q|(\bar{q}b)|b\rangle$.
For comparison, 
we also list the corresponding numbers for $\bar{B}^0$ decay
from Ref. \cite{ball} by adding up appropriate 
diquark-antidiquark decay rates.
For example, the rate of $b \rightarrow (cd)\bar{u}$ would correspond 
to the sum of the rates of 
$\bar{B}^0 \rightarrow (cd)(\overline{ud}), (cd)(\overline{ud})^*$. 
Unlike the $b \rightarrow \Dd \bar{q}$ case,
calculating the latter not only involve diquark decay constants and masses,
it also depends on $\bar B$ meson to diquark form factors. 
The diquark decay constants we use are 
 \cite{ball,decayconst},	
\beqa
g_{ud},\ g_{us}& = & 0.179, \ 0.215 \,\, {\rm GeV}^2,
   \nonumber \\
g_{cd} \cong g_{cs} & \cong & 0.35 \,\, {\rm GeV}^2,	 
\eeqa
and the diquark masses are:
\beq
m_{ud},\ m_{us},\ m_{cd},\ m_{cs} =  0.5,\ 0.7,\ 1.7,\ 2.0 \,\, {\rm GeV}.
\eeq

\begin{table}[t!]
\caption{Estimate of inclusive baryonic branching ratios.
The line separates charmed vs.\ charmless final states.}
\begin{tabular}{lcc} 
& $b \rightarrow \Dd \bar{q}$ & $B \rightarrow \Dd \overline{\Dd'}$ {\cite{ball}}
 \\ \hline 
$b \rightarrow (cd)\bar{u}$& $2.2\times 10^{-2}$ & $2.0\times 10^{-3}$ \\
$b \rightarrow (cs)\bar{c}$ & $2.4\times 10^{-2}$ & $5.8\times 10^{-3}$ \\              
$b \rightarrow (cs)\bar{u}$ & $1.1\times 10^{-3}$    &  $2.0\times 10^{-4}$  \\
$b \rightarrow (cd)\bar{c}$ & $1.3\times 10^{-3}$  & $3.0\times 10^{-4}$  \\
\hline    
$b \rightarrow (ud)\bar{u}$ & $4.8\times 10^{-5}$ & $5.2 \times 10^{-6} $  \\
$b \rightarrow (us)\bar{u}$ &  $2.1\times 10^{-5}$  & $3.9\times 10^{-7} $ \\
$b \rightarrow (ds)\bar{d}$ & $2.0\times 10^{-5}$  & $ 0  $   
\end{tabular}
\end{table}

As expected, the inclusive baryonic decays are dominated by the two charmed
modes: $b \rightarrow (cd)\bar{u}$ and $b \rightarrow (cs)\bar{c}$. 
The combined branching ratio is about $4.6 \%$. 
Adding in the rates of the smaller modes 
$b \rightarrow (cs)\bar{u}$ and $b \rightarrow (cd)\bar{c}$ gives 
a prediction for the total inclusive baryonic $B$ decay branching ratio of
\beq
{\rm Br}\,(B \rightarrow {\rm baryon + X}) = 4.8 \%. \label{baryrate}
\eeq
This is in reasonably good agreement with
the experimental result \cite{PDG}:
\beq
{\rm Br}\,(B \rightarrow {\rm baryon + X}) = 6.8 \pm 0.6 \%.
\eeq 
The minor deficit is to be expected in consideration of the possibility of
decaying into vector diquarks and other excited states as well as 
other mechanisms such as current produced baryons \cite{three1,three2}.
Though Ref.\ \cite{ball} quotes a reasonable prediction from \cite{NY},
their approach of simply adding up the $B \rightarrow \Dd\overline{\Dd'}$ 
rates would have given a branching ratio that is too small, $\sim 0.8 \%$.
This is an indication that $B \rightarrow 
\Dd\overline{\Dd'}$is not inclusive enough.
In fact, our discussion shows that this is rather an estimate of 
the fraction of the baryonic events where both baryons are energetic.

We make some observations before turning to exclusive modes.
For baryonic decays, the single charm channel $b \rightarrow (cd)\bar{u}$ 
has roughly the same rate as
the double charm channel $b \rightarrow (cs)\bar{c}$ 
since the decay constants $g_{cd}$ and $g_{cs}$ are equal 
and both have two body phase space.
This is different from the quark level picture for inclusive $b$ decays, 
where $b\rightarrow c \bar{c} s$ is suppressed by 
a factor of $3 - 5$ compared to $b\rightarrow c \bar{u} d$ 
because of having two massive final quarks in three body phase space. 
In Ref.\ \cite{ball}, i.e. for $B \rightarrow \Dd\overline{\Dd'}$, 
the difference is even more dramatic:
$b \rightarrow (cs)\bar{c}$ is more than twice $b \rightarrow (cd)\bar{u}$
because of $B \rightarrow \Dd$ form factors.
This feature of the diquark model can be tested by experiment. 
For example, one can study the inclusive decays of $\bar{B} \to \Xi_c^0 + X$ 
and $\bar{B} \to \Sigma_c^0 + X$. 
These decays arise dominantly from $b \rightarrow (cs)\bar{c}$ 
and $b \rightarrow (cd)\bar{u}$, 
with the diquark $(cs)$ or $(cd)$ picking up a $d$ quark.
The diquark model would predict
${\rm Br} (\bar{B} \to \Xi_c^0 + X) \sim {\rm Br}(\bar{B} \to \Sigma_c^0 + X)$ 
in strong contrast to expectation that
$b \to c \bar{c} s$ is less than $b \to c \bar{u} d$
by a factor of three or more.
${\rm Br}\,(\bar{B} \to \Xi^0_c X) \times {\rm Br}\,(\Xi^0_c \to \Xi^- \pi^+)$
has been measured by CLEO \cite{CLEO2}.  
While $\Sigma_c^0$ decays into $\Lambda_c \pi^-$ with 
$100\%$ branching ratio, one would need absolute measurements of
${\rm Br}\,(\Xi^0_c \to \Xi^- \pi^+)$ to perform the test.

For mesonic final states, the charmed rates are 40 to 50 times
larger than the corresponding charmless rates. 
In baryonic decays, however, $b \rightarrow (cd)\bar{u}$ is 
400 times larger 
than $b \rightarrow (ud)\bar{u}$. 
Part of the reason is that the charmed diquark decay constant is larger: 
$g_{cd} \sim 2 g_{ud}$. 
On the other hand, while penguin operators enhance charmless mesonic decays,
similar enhancement is much weaker in the baryonic modes, as we will discuss
in the next section. 
We can calculate from Table I the total inclusive
charmless baryonic decay branching ratio from the diquark picture:
\beq
{\rm Br}\,(B \rightarrow {\rm charmless \,\, baryons + X}) = 8.9 \times 10^{-5}, 
\eeq
which is relatively small.
Although this estimate is probably less reliable than Eq.\ (14),
considering the numerous possible modes 
to be discussed in detail in the next section, the largest two body decay
{\it $\bar{B}^0 \rightarrow p \bar{p}$ is likely below $10^{-6}$, 
considerably smaller than charmless mesonic decays that are 
typically of order $10^{-5}$}.

In view of the small two body branching ratios, 
it is possible that three body decays could be larger.
In a calculation analogous to that of 
$B^0 \rightarrow D^{*-} p \bar{n}$ \cite{three1}, 
it was estimated that $B^0 \rightarrow \rho p \bar{n}$ 
should be of order $10^{-5}$ \cite{three2},
hence considerably larger than two body modes.
We note that the mechanism advocated in Ref.\ \cite{three2},
that of current produced $p\bar{n}$ pair,
is not contained in the diquark model discussed here.

\section{Exclusive Decays}

As described above, two body baryonic decays proceed via 
$b\to \Dd^{(*)} \bar q$ through diquark operators. 
The diquark $\Dd^{(*)}$ captures a quark from vacuum quark pair creation 
to form a baryon.
The antiquark $\bar q$ pairs up with the spectator antiquark to form 
an anti-diquark $\overline{\Dd}^{\prime(*)}$, which then captures 
the antiquark from pair creation and becomes the antibaryon. 
Admittedly, this is not a simple process compared to meson pair
formation. We have seen that
${\cal H}_{\rm diquark}$ generates only 
scalar diquarks from $b$ decay, and hence octet baryons,
while ${\cal H}_{\rm diquark^*}$ generates only vector diquarks 
and hence decuplet baryons.
Octet and decuplet antibaryons can result from $b$ decay 
mediated by either ${\cal H}_{\rm diquark}$ or ${\cal H}_{\rm diquark^*}$. 

We find that most decay channels involve only one diquark operator.  
We shall follow Ref. \cite{ball} which calculates the matrix element of the
operators by a decomposition into four components: 
the diquark decay constant $g_{\scriptsize\textsf{\textbf{D}}}$, 
the $\bar B$ meson to antidiquark form factor 
$\langle \overline{\Dd}^\prime|(qb)|\bar B \rangle$, 
the quark pair creation wavefunction,
and the baryon wavefunction (a diquark and a quark form a baryon). 
The authors adopt a pole model to calculate the form factor,
take harmonic oscillator wavefunction in the ground state 
as the baryonic wavefunction and use nonlocal wavefunction for pair creation.
Since the latter consists of an undetermined normalization factor, 
together with other uncertainties of the four steps,
the diquark model can not be expected to predict absolute exclusive rates.
But the model may give a reasonable estimate of ratios of decay rates.

It will become clear that the penguin contributions usually  
involve the same or similar matrix elements as 
the tree contributions. 
As a result, the matrix elements calculated in Ref.\ \cite{ball} 
can be used directly in our evaluation of the penguin effects.
Since the purpose of this paper is to investigate the effect of 
penguin operators in the diquark model, we do not attemp at improving 
the calculation of amplitudes in Ref.\ \cite{ball}, 
expecting most of our conclusions to be insensitive to details.
We refer readers to Ref.\ \cite{ball} for a discussion of 
the methods of calculating the various factors.

\begin{figure}[t!]
\centerline{\DESepsf(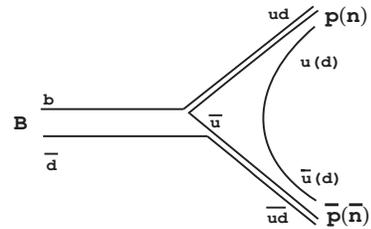 width 5.0cm)
 }
\smallskip\smallskip\smallskip
\caption{$B \rightarrow p \bar{p}, n \bar{n}$ through 
$(ud)\overline{(ud)}$. 
}
\end{figure}

\subsection{Non-strange decays $\bar{B} \rightarrow  \Bb_1 \bar{\Bb}_2$}

Let us start with the non-strange decays $\bar{B} \rightarrow \Bb_1\bar{\Bb}_2$,
such as $\bar{B}^0 \rightarrow p\bar{p}, \, n\bar{n}$. 
Taking $\bar{B}^0 \rightarrow p\bar{p}$ as an example, the decay occurs only  
through the $\bar{B}^0 \rightarrow (ud)(\bar{u}\bar{d})$ diagram, 
as shown in Fig.\ 2.
The other decay $\bar{B}^0 \rightarrow n\bar{n}$ is obtained by
replacing the $u\bar{u}$ pair
 by $d\bar{d}$.
Note that the decay through $\bar{B}^0 \rightarrow (dd)(\bar{d}\bar{d})$
 is impossible due to the antisymmetry of the constituent quark flavors 
in a scalar diquark.  
The decay rate can be written as 
\beq
 \Gamma(\bar{B}^0 \rightarrow p\bar{p}) = |{\cal A}_1|^2 \times 
|\langle p\bar{p}| \,(ub)_k (ud)_k^{\dagger}\,|\bar B\rangle|^2. 
\eeq
The constant ${\cal A}_1$, as defined in Eq.\ (\ref{a1a2})
and evaluated at the scale $\mu=m_b$, is equal to
$
4.1\times 10^{-3} \, e^{-i\gamma} + 4.5 \times 10^{-4}.   
$
The rate without penguins, as cited in Ref.\ \cite{ball}, is
\beqa
& & 
|V_{ud}^*V_{ub}(c_2-c_1)\Big|^2 \times
 |\langle p\bar{p}| \,(ub)_k (ud)_k^{\dagger}\,|\bar B\rangle|^2
    \nonumber \\
& & =\Big| 4.5\times 10^{-3} \, e^{-i\gamma} |^2 \times | 
\langle p\bar{p}| \,(ub)_k (ud)_k^{\dagger}\,|\bar B\rangle|^2. 
\eeqa
Note that the two expressions differ only in the short distance coefficients
and they share the same matrix element. 
The matrix element $\langle p\bar{p}|\,(ub)_k (ud)_k^{\dagger}\,|\bar B\rangle$ 
will be taken from Ref. \cite{ball}. 
Some interesting observations can be made even without obtaining
the absolute value.

\begin{figure}[t!]
\centerline{\DESepsf(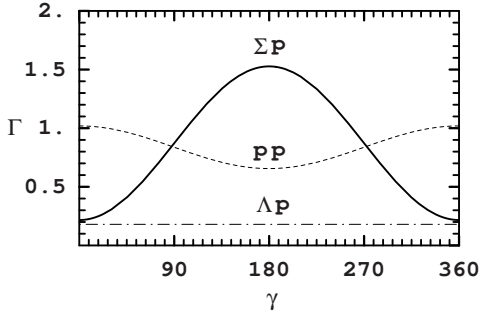 width 7.0cm)}
\smallskip\smallskip\smallskip
\caption{The $\gamma$ dependence of of 
$B\rightarrow p \bar{p}$ (dashed),  
$\Sigma^+ \bar{p}$ (solid) and $\Lambda \bar{p}$ (dot-dashed) rates in arbitrary units.
}
\end{figure}

The rate now depends on the unitarity phase angle $\gamma$ 
due to the interference of penguins with trees, 
analogous to the mesonic decays \cite{hou}. 
This dependence is shown in Fig.~3, together with the
$\gamma$ dependence of strange decays discussed later.
The penguin contribution is roughly one tenth of tree in amplitude, 
and its effect is milder compared to $B \rightarrow \pi^+\pi^-$. 
The reason is because, unlike $B \rightarrow \pi^+\pi^-$, 
which receives sizable $O_6$ contribution through chiral enhancement, 
the $O_6$ operator does not contribute in the diquark picture, 
as discussed earlier.

\begin{table}[b!]
\caption{
Relative rates of $\bar{B} \rightarrow \Bb_1 \bar{\Bb}_2$, normalized to 
$\Gamma(p \bar{p})_{\rm no \,penguin} =1$ for diquark approach [9],
and $\Gamma(p \bar{p}) =1$ in sum rule approach [8].
The angle $\gamma$ is taken to be $90^{\circ}$.
}
\begin{tabular}{lccc} 
  & This Work &  Ref. \cite{ball} & Ref. \cite{sumrule}
\\ \hline 
$ p \bar{p}$ & 0.84 & 1 & 1 \\         
$ n \bar{n}$ & 0.84 & 1 & 0.3 \\      
$ n\bar{p}$ & 0 & 0 & 0.6 \\
$p \overline{\Delta^{+}}$  & 0.28 & 0.33 &0.1   \\ 
$n \overline{\Delta^{0}}$  & 0.28 & 0.33 &    \\ 
$ p \overline{\Delta^{++}}$  & 0.63 & 0.75 &  0.25 \\ 
$ n \overline{\Delta^{+}}$  & 0.70 & 0.83 &         
\end{tabular}
\end{table}

Other $\bar{B} \rightarrow \Bb_{1} \bar{\Bb}_2$ modes exhibit similar properties.
Their rates are listed in Table II, where all rates are normalized
to the $p\bar{p}$ mode.  
Decays involving a decuplet antibaryon, 
$\bar{B} \rightarrow \Bb_{1} \bar{\Bb}^*_2$, 
is also possible, as mentioned in Sec.\ II.  
An interesting channel to search for is the mode $B^- \rightarrow n \bar{p}$.
Ref. \cite{ball} points out that this decay is impossible in the diquark picture.
This prediction is still true even after the penguin
contribution is taken into account.
None of the candidate decay diagrams such as 
the tree $B^- \rightarrow (ud)(\bar{u}\bar{u})$ and
the penguin $B^- \rightarrow (dd)(\bar{u}\bar{d})$ 
survive due to the antisymmetry of the constituent quark in a scalar diquark. 
This is very different from the sum rule calculation \cite{sumrule}, 
in which ${\rm Br}(B^- \rightarrow n\bar{p} )$ 
is about as large as ${\rm Br}(\bar{B}^0 \rightarrow p \bar{p})$.
Thus in the diquark model, one has the dynamical result of 
${\rm Br}(\bar{B}^0 \rightarrow p \bar{p}) \sim 
{\rm Br}(\bar{B}^0 \rightarrow n\bar{n})$,
but ${\rm Br}(B^+ \rightarrow p\bar{n})$ and 
${\rm Br}(B^- \rightarrow n\bar{p})$ vanish, as seen from Table II.    

The tree operators $O_{1,2}$ and penguin operators $O_{3,4}$
generate scalar diquarks and produce only 
the $\bar{B}$ decays into
an octet baryon plus an octet or decuplet antibaryon 
$\bar{B} \to \Bb \overline{\Bb}^{(*)}$. 
However, $O_{5,6}$ could generate vector diquark 
and hence $\bar{B} \to \Bb^* \overline{\Bb}^{(*)}$
is possible. 
Decays $ \bar{B}^0 \to \Delta^+ \bar{p}, \, \Delta^0 \bar{n}, \,
\Delta^+ \overline{\Delta^+}, \,\Delta^0 \overline{\Delta^0}$
would arise from the vector diquark operator 
$(u_{\sss R} b_{\sss L})^* (u_{\sss R} d_{\sss L})^{*\dagger}$,  
while $ \bar{B}^0 \to \Delta^- \overline{\Delta^-}$ and   
$ B^- \to \Delta^0 \overline{\Delta^+}, \, \Delta^- \overline{\Delta^0}$
from $(d_{\sss R} b_{\sss L})^*(d_{\sss R} d_{\sss L})^{*\dagger}$.
The amplitudes of these decays are proportional to ${\cal B}_1 \sim
1.8 \times 10^{-4}$, which is about
one twentieth of ${\cal A}_1 \sim 4.5 \times 10^{-3}$. 
The vector diquark decay constants are roughly 
the same as scalar diquark decay constants \cite{decayconst}:
\beqa
g_{ud*} &=& 0.216 \,\,{\rm GeV^2},   \nonumber \\
g_{us*} &=& 0.245 \,\,{\rm GeV^2}.
\eeqa 
Assuming that the respective form factors are also of the same order as
that for $\bar{B}^0 \to p \bar{p}$,
we expect the branching ratio of $\bar{B} \to \Bb^* \overline{\Bb}^{(*)}$
to be about  
$ 0.0025 \times {\rm Br}\,(\bar{B}^0 \to p \bar{p})$. 
The small rates of $\bar{B} \to \Bb^* \overline{\Bb}^{(*)}$ is 
a testable feature of the diquark model.
Because of further numerical uncertainties, 
this type of modes are not listed in Table~II.

\subsection{Strange decays $\bar{B} \rightarrow \Bb_s \overline{\Bb}$}

In mesonic $B$ decays, strange decays like $B \rightarrow K\pi$ 
have larger rates than nonstrange decays like $B \rightarrow \pi\pi$ 
because of penguin contributions, with 
${\rm Br}\,(B \rightarrow K^-\pi^+) \approx 1.88 \times 10^{-5}$ compared to
${\rm Br}\,(B \rightarrow \pi^-\pi^+) \approx 4.7 \times 10^{-6}$. 
One may wonder if the same could happen in the baryonic decays between 
$\bar{B} \rightarrow \Bb_{s} \bar{\Bb}$, such as  
$\bar{B}^0 \rightarrow\Lambda \bar{n},\ \Sigma^+\bar{p},\ \Sigma^0\bar{n}$, vs
$\bar{B} \rightarrow \Bb_1 \bar{\Bb}_2$ such as $\bar{B}^0 \rightarrow p\bar{p}$. 
This is indeed the case in the sum rule calculation, which gives 
${\rm Br}\,(\bar{B}^0 \rightarrow p\bar{p})=1.6\times 10^{-6}$ 
while ${\rm Br}\,(\bar{B}^0 \rightarrow\Sigma^+\bar{p})=6\times 10^{-6}$.
In the diquark caculation of Ref. \cite{ball},
${\rm Br}\,(\bar{B}^0 \rightarrow\Sigma^+\bar{p})$
is only 0.15 times ${\rm Br}\,(\bar{B}^0 \rightarrow p\bar{p})$.
However, since the penguin operators are not included in Ref. \cite{ball}, 
it is of interest to include the penguins to find the actual
prediction of the diquark picture.

\begin{figure}[b!]
\centerline{\DESepsf(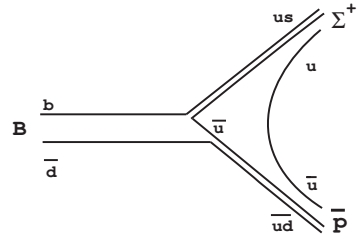 width 5.0cm)}
\smallskip\smallskip\smallskip
\caption{$\bar{B}^0\rightarrow \Sigma^+ \bar{p}$ .
}
\end{figure}

To include penguins, 
we proceed just like in the discussion of non-strange decays.
Taking $\bar{B}^0 \rightarrow \Sigma^+\bar{p}$ as an example,
only one diquark diagram, Fig.\ 4, through $\bar{B}^0 \rightarrow
(us)(\bar{u}\bar{d})$ will contribute,
\beq
\Gamma(\bar{B}^0\rightarrow \Sigma^+\bar{p}) = |{\cal A}_2|^2 \times 
 | \langle \Sigma\bar{p}| \,(ub)_k (us)_k^{\dagger}\,|\bar B\rangle|^2,
\eeq
The constant ${\cal A}_2$ as defined in Eq.\ (\ref{a1a2}), 
evaluated at the scale $m_b$, is equal to
$1.0\times 10^{-3} \, e^{-i\gamma} -2.2 \times 10^{-3} $. 
The rate without penguins is 
\beqa
& &
| V_{us}^*V_{ub}(c_2-c_1)|^2 \times  
 | \langle \Sigma\bar{p}|\, (ub)_k (us)_k^{\dagger}\,|B\rangle|^2   \nonumber \\
&  & = | 1.0\times 10^{-3} \, e^{-i\gamma} |^2 \times 
 | \langle \Sigma\bar{p}|\,(ub)_k (us)_k^{\dagger}\,|B\rangle|^2.
\eeqa
Again the two expressions share the same matrix element, 
which we take from Ref. \cite{ball} and find that
\beq
|\langle \Sigma\bar{p}| \,(ub)_k  (us)_k^{\dagger}\,|B\rangle|^2 \simeq
 3.0 \times  |\langle p\bar{p}| \,(ub)_k (ud)_k^{\dagger}\,|B\rangle|^2.
\eeq

The expression for ${\cal A}_2$ indicates that the contribution from penguins
is almost twice as large as the tree contribution in amplitude
and can not be ignored. 
The actual branching ratio of $\bar{B}^0 \rightarrow \Sigma^+\bar{p}$ 
will depend on the angle $\gamma$ (see Fig.\ 3).
The penguin operators do enhance the rate significantly 
and ${\rm Br}\,(\bar{B}^0\rightarrow\Sigma^+\bar{p})$
is larger than ${\rm Br} \,(\bar{B}^0\rightarrow p\bar{p})$
for $\gamma>90^{\circ}$. 
However, the effects are 
milder than in the mesonic decays. 
${\rm Br}\,(\bar{B}^0\rightarrow\Sigma^+\bar{p})$
is at most twice ${\rm Br} \,(\bar{B}^0\rightarrow p\bar{p})$, when
$\gamma=180^{\circ}$. 
This is largely because 
the operators $O_{5,6}$ do not contribute to the same final state,
unlike the role of chiral enhancement and constructive interference
(between $O_4$ and $O_6$) in $B \rightarrow K\pi$.
Analogous to $B \rightarrow K \pi$ and $B \rightarrow \pi \pi$, 
large $\gamma$ will enhance ${\rm Br}\,(\bar{B}^0\rightarrow\Sigma^+\bar{p})$ 
but decrease ${\rm Br}\,(\bar{B}^0 \rightarrow p\bar{p})$. 
Similar results hold for the decays 
$\bar{B}^0\rightarrow\Lambda \bar{n},\ \Sigma^0 \bar{n}$.

\begin{figure}[t!]
\centerline{\DESepsf(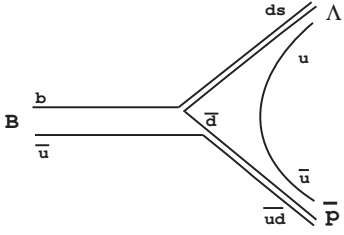 width 5.0cm)}
%
\smallskip\smallskip\smallskip
\caption{
$B^- \rightarrow \Lambda \bar{p}$ through  
$(ds)\overline{(ud)}$. 
}
\end{figure}

There are pure penguin contributions that were
not given in Ref.\ \cite{ball}.
The modes $B^-\to\Lambda \bar{p}$, $\Sigma^0 \bar{p}$ are two examples.
Nonzero tree contribution would require the diquark decay channel 
$B^- \rightarrow (us)(\bar{u}\bar{u})$, 
which is impossible due to the antisymmetry of constituent. 
However, these modes can be generated through the diquark operator 
$(db)_k \, (ds)_k^{\dagger}$, which arises only from penguin operators,
as shown in Fig. 5.
This diagram is not calculated in Ref.\ \cite{ball}, but is identical to 
$\bar{B}^0\rightarrow\Lambda \bar{n},\ \Sigma^0 \bar{n}$, respectively, 
after an isospin transformation $u \leftrightarrow d$.
Hence the rate for $B^- \rightarrow\Lambda \bar{p}$ is given by
\beq
\Gamma(B\rightarrow\Lambda\bar{p}) =  |{\cal A}_4|^2 \times  
| \langle \Lambda\bar{n}| (ub)_k  (us)_k^{\dagger}|\bar B\rangle|^2.
\eeq
The coefficient $A_4$ is equal to $-2.2 \times 10^{-3}$.
Since there is no tree-penguin interference, the rates for 
$B^-\rightarrow\Lambda \bar{p}, \, \Sigma^0 \bar{p}$ are 
independent of the angle $\gamma$, just like $B\rightarrow K^0 \pi^0$

\begin{table}[t!]
\caption{The relative rates of $\bar{B} \rightarrow \Bb_s \bar{\Bb}$
normalized as in Table II.
The angle $\gamma$ is taken to be $90^{\circ}$. }
\begin{tabular}{lccc} 
 & This Work   & Ref. \cite{ball} & Ref. \cite{sumrule}
\\ \hline 
$p \bar{p}$ & 0.84 & 1 & 1 \\                
$ \Sigma^{+} \bar{p}$ & $0.88$ & $0.15$ & 7.5  \\ 
$\Sigma^0 \bar{n}$ & $ 0.21$ & $0.037$ &  \\ 
$ \Sigma^- \bar{n}$ & $0.72$ & 0 &  \\    
$ \Sigma^0 \bar{p}$ & $0.18$ & 0 & 3.8 \\ 
$ \Lambda \bar{n}$ & $0.21$ & $0.037$ &  \\
$ \Lambda \bar{p}$ & $0.18$ & 0 & $< 3.8$ \\      
$  \Sigma^+\overline{\Delta^{+}}$ &  0.57 & $0.10$  & 7.5   \\
$   \Sigma^0\overline{\Delta^{0}}$ &  0.17 & $0.030$  & \\
$ \Sigma^+ \overline{\Delta^{++}}$ & $1.1 $ & 0.2 & 7.5 \\ 
$ \Sigma^0 \overline{\Delta^{+}}$ & $0.080 $ & 0.014 & \\  
$   \Lambda\overline{\Delta^{0}}$ &  0.086 & $0.015$  &  \\
$ \Lambda \overline{\Delta^{+}}$ & $0.023 $ & 0.004 & \\
\end{tabular}
\end{table}

 The branching ratios of the modes $\bar{B} \rightarrow \Bb_{s} \bar{\Bb}$ 
are listed in Table III.
For comparison, the sum rule \cite{sumrule} results are also listed,
which gives a strikingly different pattern. 
$B^- \to \Sigma^+ \overline{\Delta^{++}}$
and $\bar{B}^0 \to \Sigma^+ \overline{\Delta^+}$
have close or larger rates than 
$\bar{B}^0 \to \Sigma^+ \bar{p}$.   
The $\gamma$ dependence of their rates are identical 
to that of $\Sigma^+ \bar{p}$.

The $O_{5,6}$ operators could generate 
$\bar{B} \to \Bb_s^* \overline{\Bb}^{(*)}$ via vector diquarks.
The decays $\bar{B}^0 \to \Sigma^{*+} \bar{p}$, $\Sigma^{*0} \bar{n}$,
$\Sigma^{*+} \overline{\Delta^+}$, $\Sigma^{*0} \overline{\Delta^0}$, 
$\Sigma^{*+} \overline{\Delta^{++}}$ 
and $B^- \to \Sigma^{*0} \overline{p}$, $\Sigma^{*+} \overline{\Delta^{++}}$
can arise from the vector diquark operator 
$(u_{R} b_{L})^* (u_{R} s_{L})^{*\dagger}$,  
while $\bar{B}^0 \to \Sigma^{*0} \overline{\Delta^0}$,
$\Sigma^{*-} \overline{\Delta^-}$ and   
$B^- \to \Sigma^{*-} \overline{n}$, $\Sigma^{*-} \overline{\Delta^0}$,
$\Sigma^{*0} \overline{\Delta^+}$
from $(d_{R} b_{L})^*(d_{R} s_{L})^{*\dagger}$.
The amplitude of these decays are proportional to 
${\cal B}_2 \sim 9.0 \times 10^{-4}$, which is about
one half of ${\cal A}_4$. 
As described above,
${\rm Br}(\bar{B}^0 \to \Lambda \bar{p})$ is proportional to ${\cal A}_4^2$.
Assuming that the form factors are of the same order,
we expect the rates of $\bar{B} \to \Bb_s^* \overline{\Bb}^{(*)}$ to be roughly 
$ 1/4 \times {\rm Br}\,(\bar{B}^0 \to \Lambda \bar{p}) \sim 0.03 \times
{\rm Br}\,(\bar{B}^0 \to p \bar{p})$. 
This is still an order magnitude smaller than the typical $\bar{B}$
decays to an octet baryon plus an octet or decuplet antibaryon. 
Again, since further uncertainties are invloved, the 
$\bar{B} \to \Bb_s^* \overline{\Bb}^{(*)}$ modes are not listed in Table III.

\subsection{$\bar{B} \rightarrow \Bb_s \bar{\Bb}_s$}

This category includes decays like
$\bar{B}^0 \rightarrow \Lambda \bar{\Lambda}$, 
$\Xi^0 \bar{\Lambda}$, etc. 
The relative rates of these modes are listed in Table IV.
The mode $\bar{B}^0 \rightarrow \Xi^0 \bar{\Lambda}$ is more
straightforward since only one diquark diagram,
$\bar{B}^0 \rightarrow (us)(\bar{u}\bar{d})$, is involved. 
The enhancement effects from penguin is very similar to 
$\bar{B}^0\rightarrow \Sigma^+\bar{p}$.

\begin{figure}[t!]
\centerline{\DESepsf(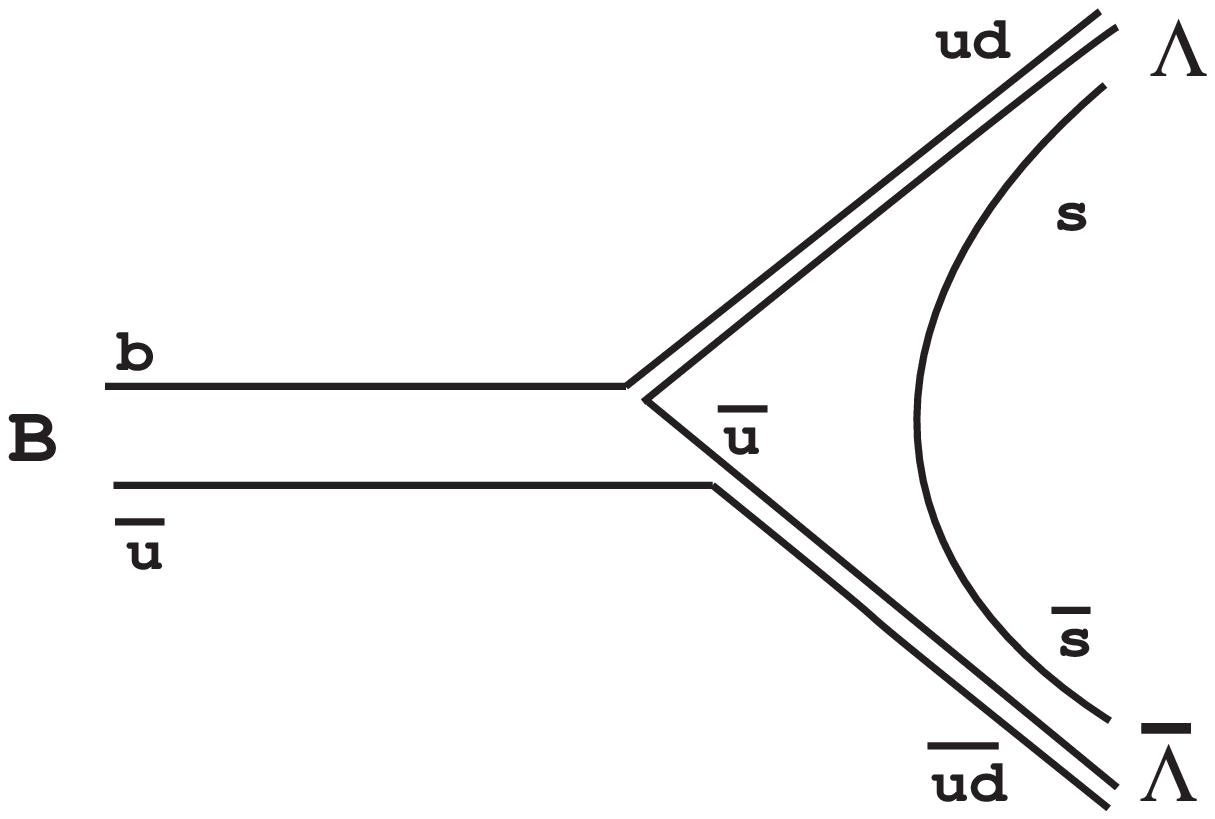 width 4.4cm)
\DESepsf(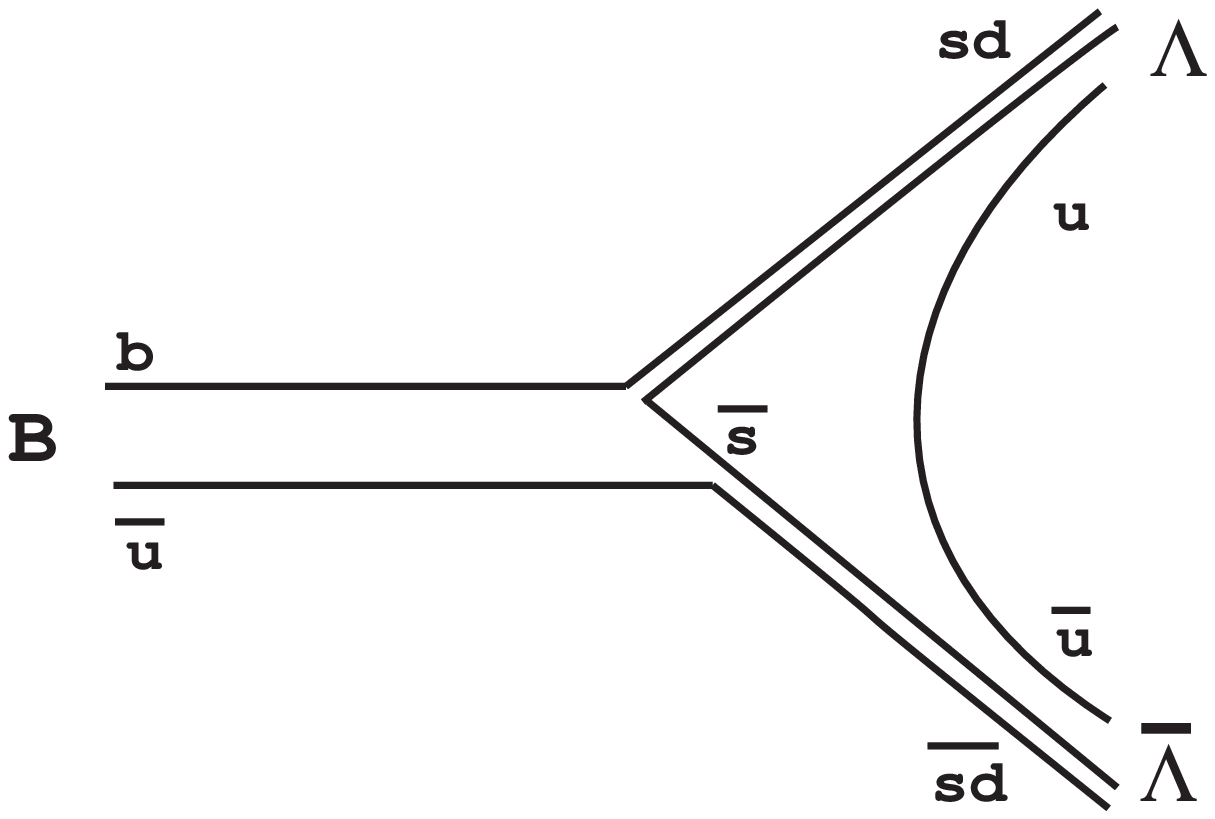 width 4.4cm)}
\smallskip\smallskip\smallskip
\caption{
$\bar{B}^0\rightarrow \Lambda \bar{\Lambda}$ through 
(a) $(ud)\overline{(ud)}$ and (b) $(sd)\overline{(sd)}$. 
}
\end{figure}

$\bar{B}^0 \rightarrow \Lambda \bar{\Lambda}$ decay can arise from two
diquark diagrams and hence is more complicated.
The tree contribution is through $\bar{B}^0 \rightarrow (ud)(\bar{u}\bar{d})$  
decay plus an $s\bar{s}$ pair creation (Fig.\ 6(a)).
The penguin operators will enhance the Wilson coefficient in this diagram 
just like in the case of $\bar{B}^0 \rightarrow p \bar{p}$.
However, there is one more contribution from the penguin operator: 
through $\bar{B}^0 \rightarrow (sd)(\bar{s}\bar{d})$   
with $u\bar{u}$ pair creation (Fig.\ 6(b)).
We shall argue that the penguin contribution is smaller than the tree.

The short distance coefficient for penguins, 
${\cal A}_3=3.5\times 10^{-4}$, is smaller than one tenth of the
tree $V_{ud}^*V_{ub}(c_2-c_1) =4.5 \times 10^{-3}$.
The matrix element for Fig.\ 6(b) is basically the same as in Fig.\ 6(a),
except replacing $s\bar{s}$ pair creation by $u\bar{u}$.  
To estimate the effect of this replacement, 
we can compare relative rates of $(\bar{B}^0 \rightarrow p \bar{p})$ 
vs $\bar{B}^0 \rightarrow \Lambda \bar{\Lambda}$ 
calculated in Ref.\ \cite{ball}, which is about $2.6 : 1$.
Since the only difference between them at tree level
is just in the pair creation, fixes the relative weight of 
$s\bar{s}$ pair creation vs $u\bar{u}$. 
This, we expect the penguin contribution overall to be one fifth of the tree.
We therefore ignore the penguin contribution
in our reporting of $\Lambda \bar{\Lambda}$ rates in Table IV. 
To obtain the actual numerical value and $\gamma$ dependence
in the future, one would have to evaluate Fig.~6(b) .  

\begin{table}[b!]
\caption{The relative rate of $B \rightarrow {\rm \bf B_s \bar{B}_s}$. 
Diquark rates are normalized  so that $\Gamma(p \bar{p})_{\rm no \,penguin} =1$.
Sum rule rates are normalized so that $\Gamma(p \bar{p})_{\rm sum \,rule} =1$.
The angle $\gamma$ is assumed to be $90^{\circ}$. }
\begin{tabular}{lccc} 
& Rate  & Ref. \cite{ball} & Ref. \cite{sumrule}
\\ \hline 
$p \bar{p}$ & 0.84 & 1 & 1 \\                
$ \Lambda \bar{\Lambda}$ & $0.38$ & $0.39$ &  \\
$ \Xi^0 \bar{\Lambda}$ & $ 0.34$ & $0.059$ &  \\ 
$\Lambda \overline{\Sigma^*}^0$ & $0.068$ & $0.082$ &   \\ 
$ \Xi^0 \overline{\Sigma^*}^0$& $0.11$ & 0.02 &  \\      
$ \Lambda \overline{\Sigma^{*+}}$ & $0.068$ & $0.082$ &   \\ 
$ \Xi^0 \overline{\Sigma^{*+}}$& $0.11$ & 0.02 &  
\end{tabular}
\end{table}

\section{Discussionand Conclusion}

In this paper, we have discussed the effects of penguin operators on 
two body baryonic $B$ decays in the diquark picture. 
We point out that the penguin operators $O_{3,4}$ can also be
transformed into operators with diquark fields and 
hence the calculation of their contributions is very similar to
that of the tree operators. 
On the other hand, $O_{5,6}$ do not generate scalar diquark operators,
indicating that their contribution to $\bar{B}$ decays into 
an octet baryon is small.  
As a result, penguin operators will enhance significantly 
$\bar{B}\rightarrow \Bb_s \bar{\Bb}$.
Though the effects may not be large enough to reverse the general relative order of 
$\bar{B}\rightarrow \Bb \bar{\Bb}$ and $\bar{B}\rightarrow \Bb_s \bar{\Bb}$   
as in the mesonic decays, some modes do have comparable rates.  
For example, $\bar{B}^0 \rightarrow \Sigma^+ \bar{p}$, 
after penguin  enhancement, is
larger than $\bar{B}^0 \rightarrow p\bar{p}$
for $\gamma > 90^{\circ}$.
The $B^- \to \Sigma^+\overline{\Delta^{++}}$ mode
is even of order 30\% larger than $\bar{B}^0 \rightarrow \Sigma^+ \bar{p}$.

The penguin operators $O_{5,6}$ could generate $\bar{B}$ decays
to a decuplet baryon.
These channels were predicted to vanish in Ref.\ \cite{ball}
since the tree operators $O_{1,2}$ only generate scalar diquarks.
However, $(V+A)\times (V-A)$ type penguin operators $O_{5,6,7,8}$ could
generate vector diquarks after Fierz transformation, 
which could form decuplet baryons as the final product.
Their rates are nevertheless small.
We estimate the non-strange and strange decays
${\rm Br} \,(\bar{B} \to \Bb^* \bar{\Bb}^{(*)})$,
${\rm Br}\,(\bar{B} \to \Bb_s^* \bar{\Bb}^{(*)})$ are about 
$0.25 \%$, 3\% of  ${\rm Br} \, (\bar{B}^0 \to p \bar{p})$,
respectively.

The diquark model calculation of the exclusive rates depends on 
the pair creation model with an undetermined normalization factor. 
Hence, absolute rates can not be obtained.
However, as a result of duality, the inclusive rates is independent of 
pair creation model. 
We estimate the inclusive rate for baryonic decays
by computing the rate of $ b \to \Dd \bar{q}$. 
This calculation relies only on the assumption of the diquark model and the
values of the diquark decay constants, without further
dynamical assumption about form factors.
The total rate we get is very close to the experiment result,
indicating that the diquark model is a reasonable
picture for baryon production.
Actually, the theoretical value is somewhat smaller, 
leaving some room for other mechanisms.

Since the inclusive  prediction relies only on the decay constants 
in the diquark model, the agreement also indicates that 
the values of $g_{cd}$ and $g_{cs}$ used are reasonable. 
Ratio of exclusive decay rates can further check 
the values of $g_{ud}$ and $g_{us}$.
For example, the modes $\bar{B}^0 \rightarrow p \, \overline{p}$ and 
$\bar{B}^0 \rightarrow \Sigma_c^+ \overline{p}$
have an identical $\bar B$ to antidiquark form factor,
and they differ only in the diquark decay constants and CKM factors. 
Assuming that the transition form factor of 
$B$ to the antidiquark $\overline{(ud)}$ 
is not very sensitive to the momentum transfer, it will cancel 
in the ratio of their rates.
Similar argument applies for the pair creation wavefunction.
The ratio can be written as :
\beq
\frac{\Gamma(\bar{B}^0 \rightarrow p \overline{p})}
{\Gamma(\bar{B}^0 \rightarrow \Sigma_c^+  \, \overline{p})} = 
\left|\frac{V_{ub}}{V_{cb}}\right|^2 \times
\left(\frac{g_{ud}}{g_{cd}}\right)^2,
\eeq
allowing one to in principle test diquark decay constant ratios.
Likewise, the ratio of $\bar{B}^0 \rightarrow \Sigma^+ \overline{\Lambda_c^+}$
to $\bar{B}^0 \rightarrow \Lambda_c^+ \overline{\Lambda_c^+}$ 
could test $g_{us}/g_{cd}$.

It should be clear from the previous sections that the diquark picture gives
rather different predictions from the sum rule calculation. 
Most significantly, we note that $\bar{B} \rightarrow \Bb_1\bar{\Bb}_2$ 
is suppressed compared to $\bar{B}\rightarrow \Bb_c \bar{\Bb},\ 
\Bb_c \bar{\Bb_c}$ and $\Bb_s \bar{\Bb}$ in sum rule treatment \cite{sumrule}.
The reason is that, as the sum rule authors claim, 
quark pair creation is mainly a soft process 
instead of a hard one like in the diquark non-local pair creation model. 
A soft process would favor heavier quarks carrying larger momentum 
 in the final product to pick up soft quarks from the vacuum. 
In the sum rule calculation, therefore, 
the amplitude for producing an additional quark from the vacuum
 is of order 1 in $\bar{B}\rightarrow \Bb_c \bar{\Bb}$ and 
$\Bb_c \bar{\Bb_c}$ but suppessed in $\bar{B}\rightarrow \Bb_1\bar{\Bb}_2$.
Such effects are much less pronounced in the diquark model. As a result, 
$\bar{B}\rightarrow \Bb_s \bar{\Bb}$ typically is still smaller than 
$\bar{B}\rightarrow \Bb_1\bar{\Bb}_2 $, 
even after penguins are taken into account.

Another feature of the diquark model is that 
several decay modes are missing due to 
the antisymmetry of the constituent quark flavor in a scalar diquark. 
For example, there is no $B^- \to n\bar{p}$ while 
$\bar{B}^0 \to p\bar{p}$ and $n\bar{n}$ have the same rates. 
The modes $B^- \rightarrow \Lambda \bar{p},\ \Sigma^0 \bar{p}$ 
are pure penguins and are smaller than $\bar{B}^0 \rightarrow p\bar{p}$. 
The sum rule approach predicts 
$B^- \rightarrow n\bar{p},\ \Lambda\bar{p},\ \Sigma^0 \bar{p}$ are
of the same order as $\bar{B}^0 \rightarrow p\bar{p}$. 
The decays arising from the penguin operator
$(\bar ss)_{V \pm A}(\bar sb)_{V-A}$ such as the novel one
$\bar{B}^0 \rightarrow \Omega \bar{\Xi^-}$ are 
supposedly possible in sum rule calculation
(though it is not mentioned in \cite{sumrule}). 
However the above penguin operator
$(\bar ss)_{V \pm A}(\bar sb)_{V-A}$ does not 
have a scalar diquark component and thus such decays should be suppressed.  
The authors of \cite{ball} estimate that these decays, 
forbidden by the scalar diquark model, 
should be suppressed by at least a factor of 3. 
The predictions emerging from the two pictures are, anyway, 
different enough to be tested in the near future 
by experimantal observation of $B$ meson baryonic decays.  

The path to observation and especially understanding charmless
baryonic $B$ decays would be a long and winding one.

\vskip 0.3cm  
\noindent{\bf Acknowledgement}.\ \  
This work is supported in part by  
the National Science Council of R.O.C.  
under Grants NSC-89-2112-M-002-063 and NSC-90-2112-M-003-010,  
the BCP Topical Program of NCTS, and the MOE CosPA Project.

\end{document}